\documentclass[CRPHYS,Unicode,manuscript]{cedram}

\renewcommand{\vec}[1]{{\mathbf #1}}

\usepackage{xcolor}

\usepackage{soulutf8}

\usepackage{amssymb}

\title{Photonic topological Anderson insulator in a two-dimensional atomic lattice}

\author{\firstname{Sergey} \middlename{E.} \lastname{Skipetrov}\CDRorcid{0000-0001-6186-1929}\IsCorresp
}
\address{Univ. Grenoble Alpes, CNRS, LPMMC, 38000 Grenoble, France}
\email[S.E. Skipetrov]{Sergey.Skipetrov@lpmmc.cnrs.fr}
\author{\firstname{Pierre}  \lastname{Wulles}\CDRorcid{0000-0002-0810-3464}}
\addressSameAs{1}{Univ. Grenoble Alpes, CNRS, LPMMC, 38000 Grenoble, France}
\email[P. Wulles]{Pierre.Wulles@lpmmc.cnrs.fr}

\thanks{This work was funded by the Agence Nationale de la Recherche (Grant No. ANR-20-CE30-0003 LOLITOP)} 

\CDRGrant[ANR]{ANR-20-CE30-0003}


\begin{abstract} 
Disorder in atomic positions can induce a topologically nontrivial phase---topological Anderson insulator (TAI)---for transverse electric optical quasimodes of a two-dimensional honeycomb lattice of immobile atoms. TAI requires both time-reversal and inversion symmetries to be broken to similar extents. It is characterized by a nonzero topological invariant, a reduced density of states and spatially localized quasimodes in the bulk, as well as propagating edge states. A transition from TAI to the topological insulator (TI) phase can take place at a constant value of the topological invariant, showing that TAI and TI represent the same topological phase.

\end{abstract}

\begin{document}

\maketitle

\section{Introduction}

Interaction of light with atomic vapors is a fascinating research field full of surprises \cite{cohen98}. On the one hand, atoms scatter light thus affecting its propagation whereas on the other hand, light can change the internal state of an atom and exert a mechanical force on it \cite{cohen11}. The most fascinating example of the practical use of light-atom interaction is probably the cooling of atomic ensembles to very low temperatures and reaching the thermodynamic phase of a Bose-Einstein condensate (BEC) in which the majority of atoms are in the same quantum state \cite{cohen11,cornell02}. The interaction of laser-cooled atoms with light has become a research field on its own.  Although its physics is, in general, very complex, regimes can be identified when only one of the components---light or atoms---acts on the other whereas there is no back action. One can, for example, consider atoms evolving in an optical potential due to external lasers without the potential being affected by the atoms in any way. Cold atomic systems, mainly BECs, in carefully engineered or random optical potentials have recently allowed for studying several intriguing physical phenomena initially proposed in the context of condensed matter physics: Anderson localization \cite{billy08,jendr12}, Mott-Hubbard transition and other many-body phenomena \cite{bloch08}, as well as phenomena related to the nontrivial topology of energy bands \cite{cooper19}.  Under different conditions, it is possible to achieve a situation in which light is scattered by atoms with the latter staying immobile in space to a good approximation \cite{labeyrie99,corman17}. In this context, attempts to observe Anderson localization of light are underway \cite{kaiser09}, even though they have to face additional difficulties due to the vectorial character of light  \cite{skip14,tiggelen21}.

Topology is a branch of mathematics that revealed its full power for understanding physical phenomena after the discovery of the quantum Hall effect (QHE) in 1980 \cite{klitzing80,klitzing20}. It turns out that the precise quantization of Hall conductance in QHE finds a simple explanation in terms of topology of electronic bands \cite{thouless82,simon83}. Since that time, physicists realized that topology can provide a useful perspective in various domains ranging from the physics of cold atoms \cite{cooper19} to photonics \cite{lu14,ozawa19} and Earth sciences \cite{delpace17}. Moreover, the robustness of topological phenomena with respect to moderate disorder and perturbations makes them attractive for applications in devices where defects and various kinds of imperfections are unavoidable (e.g., in nanoelectronics). This robustness was the main reason to discuss the impact of disorder in the context of topological physics until 2009 when a possibility  {was} discovered for disorder to play a constructive role and to {\it induce} nontrivial topological phases instead of destroying them, leading to a topological phase dubbed the ``topological Anderson insulator'' (TAI) \cite{li09}.
 {First discovered in a system described by a tight-binding Hamiltonian belonging to the symmetry class AII and uncorrelated on-site disorder \mbox{\cite{li09}}, TAI phase has been later demonstrated to exist in systems of other symmetry classes and with different types of disorder (including amorphous systems) as well \mbox{\cite{xing11,orth16,agarwala17}}.}
Whereas a topological insulator (TI) features edge states with frequencies inside a band gap of the infinite system \cite{bernevig13}, TAI is supposed to exhibit the same behavior but only when disorder is introduced in the system, which explains the reference to Anderson in its name.
Spatially localized states  arise in the
 {bulk, leading to a mobility gap in the spectrum.} The concept of TAI has been extended to optical systems \cite{liu17,stutzer18,liu20,zhou20,cui22}. 
   
In this paper we explore the phenomenon of TAI in a system of cold, immobile atoms interacting via the electromagnetic field. Our atomic system is obtained from a honeycomb lattice of atoms studied in Ref.\ \cite{perczel17} by introducing disorder in atomic positions. The possibility of obtaining a TAI in such a system has already been identified in our recent publication \cite{skip22}. Here we provide its detailed characterization, identify the range of parameters suitable for its experimental observation, and discuss its interplay with Anderson localization. Before presenting our results, we would like to attract reader's attention to the essential differences between solid-state (electronic) and optical systems that make a photonic TAI different from its electronic counterpart. First, in contrast to electrons, photons do not have a charge and thus an external magnetic field can only act on the photons indirectly, by modifying the properties of the medium in which photons propagate. Second, optical systems are often open and thus non-Hermitian. Third, photons have a polarization that is not simply an additional degree of freedom but can crucially modify the impact  {of} disorder on photon propagation by, e.g., preventing Anderson localization \cite{skip14,tiggelen21}. All these particularities will play their role in the atomic system that we consider below.

\section{The model}
\label{model}

\begin{figure}[tbp]
\includegraphics[width=\textwidth]{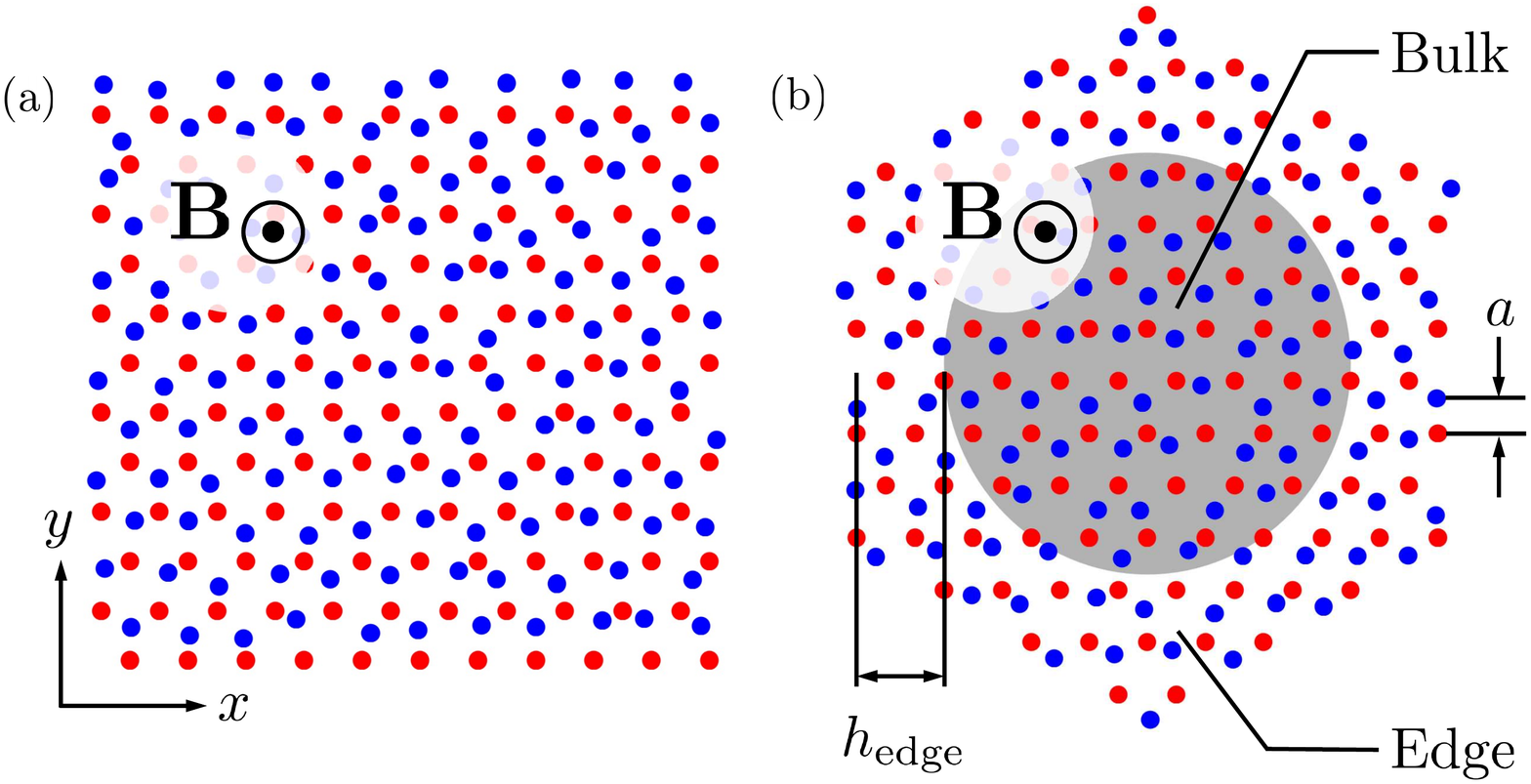}
\vspace{-3cm}
\caption{\label{figlattices}
Schematic representations of disordered honeycomb atomic lattices studied in this work
 {(for the purpose of illustration, the lattices shown are much smaller than the actual ones used in the calculations)}. Atoms $A$ shown in red form a perfect triangular lattice with a lattice spacing $a \sqrt{3}$. Atoms $B$ shown in blue are displaced from their positions in the second triangular lattice (obtained by translating the first lattice by a distance $a$ along the vertical axis $y$) by random distances $\Delta r_m$ uniformly distributed between 0 and $Wa$ in random directions ($W = 0.4$ in the figure). A constant, spatially uniform magnetic field $\vec{B}$ is applied perpendicular to the plane of the atomic lattice. (a) A rectangular lattice used for the calculation of Bott index. The lattice used for the calculations in this work consists of $N = 2244$ atoms and is as close as possible to a square.  (b) A lattice having the shape of a hexagon with armchair edges that do not support edge states in the absence of magnetic filed (i.e., for $\Delta_{\vec{B}} = 0$). The lattice used for the calculations in this work consists of $N = 4326$ atoms. The edge of the lattice is defined by $h_{\text{edge}} = (N_{\text{edge}}-1) a \sqrt{3}/2$ and the remaining grey part is referred to as ``bulk''. We use $N_{\text{edge}} = 4$ in this work.}
\end{figure}

We consider a two-dimensional (2D), planar lattice of $N/2$  unit cells composed of two atoms $A$ and $B$ that may be different (see Fig.\ \ref{figlattices}). The lattice is embedded in the three-dimensional (3D) infinite free space. We denote resonance frequencies of atoms $A$ and $B$  by $\omega_A$ and $\omega_B$, respectively; the natural decay rate of the excited state is assumed to be the same and equal to $\Gamma_0$ for all atoms. Each atom has a nondegenerate ground state of angular momentum $\vec{J}_g = 0$ and a  {three} degenerate excited states of $\vec{J}_e = 1$  {with projections $J_z = 0, \pm 1$ of $\vec{J}_e$ on the quantization axis $z$}. An external magnetic field $\vec{B} = \{ 0, 0, B_z \}$ perpendicular to the atomic plane $xy$ induces a dimensionless Zeeman shift  $\Delta_{\vec{B}} = \mu_B B_z/\Gamma_0$ ($\mu_B$ is the Bohr magneton)
 {of energies of $J_z = \pm 1$ states in opposite directions.}
We assume that atoms $A$ form a triangular lattice with a lattice spacing $a \sqrt{3}$ and denote their positions by $\{ \vec{r}_m \} = \{ x_m, y_m \}$, $m = 1, \ldots, N/2$. The atoms of the sublattice $B$ are located at $\{ \vec{r}_m + \vec{a}_m \}$, where $\vec{a}_m$ denotes the position of the atom $B$ with respect to the atom $A$ of the same unit cell $m$. A honeycomb lattice is obtained when all $\vec{a}_m$ are the same and directed along the $y$ axis: $\vec{a}_m = \vec{a} = \{0, a \}$. We will consider a situation in which $\vec{a}_m$ are independent identically distributed random vectors with a mean value $\langle \vec{a}_m \rangle = \vec{a}$. This corresponds to a randomized honeycomb lattice. The parameter $\Delta_{\vec{B}}$ describes the degree of the time-reversal (TR) symmetry breakdown in the atomic system, whereas $\Delta_{AB} = (\omega_B - \omega_A) /2\Gamma_0$ measures the breakdown of the inversion symmetry (i.e., the symmetry with respect to the exchange of atoms $A$ and $B$). Finally, the disorder in atomic positions leads to the breakdown of the discrete translational symmetry. The interplay of these three symmetry breakdowns is at the heart of the phenomenon of TAI studied in this work.   

 {Consider now} coupling between the atoms due to quasiresonant electromagnetic waves with the electric field in the atomic plane---transverse electric (TE) modes.  {These modes will interact only with atomic transitions from the ground state to the excited states with $J_z = \pm 1$ whereas the transition to $J_z = 0$ state will remain idle. Taking into account only these two transitions, we write} the effective Hamiltonian $\hat{H}$ of the system  {(in units of $\hbar \Gamma_0$)} as a $2N \times 2N$  {non-Hermitian} matrix
\begin{eqnarray}
{\hat H} &=& 
\begin{bmatrix}
{\hat H}_{11} & {\hat H}_{12} & \cdots & {\hat H}_{1(N/2)} \\
{\hat H}_{21} & {\hat H}_{22} & \cdots&  {\hat H}_{2(N/2)} \\
\cdots  & \cdots & \cdots & \cdots \\
{\hat H}_{(N/2)1} & {\hat H}_{(N/2)2} & \cdots & {\hat H}_{(N/2)(N/2)}
\end{bmatrix}
\label{fullh}
\end{eqnarray}
composed of $4 \times 4$ blocks \cite{perczel17,skip22}
\begin{eqnarray}
{\hat H}_{mn} &=& \delta_{mn} 
\left\{
\begin{bmatrix}
i \mathbb{1} & \hat{G}(-\vec{a}_m)\\
\hat{G}(\vec{a}_m) & i \mathbb{1}
\end{bmatrix}
+  2 \Delta_{AB}
\begin{bmatrix}
\mathbb{1} & 0\\
0 & -\mathbb{1}
\end{bmatrix}
+ 2 \Delta_{\vec{B}}
\begin{bmatrix}
{\hat \sigma}_z & 0\\
0 &{\hat \sigma}_z
\end{bmatrix}
\right\}
\nonumber \\ 
&+& (1 - \delta_{mn}) 
\begin{bmatrix}
\hat{G}(\vec{r}_m - \vec{r}_n) & \hat{G}(\vec{r}_m - \vec{r}_n - \vec{a}_n)\\
\hat{G}(\vec{r}_m - \vec{r}_n + \vec{a}_m) & \hat{G}(\vec{r}_m - \vec{r}_n + \vec{a}_m - \vec{a}_n)
\end{bmatrix}\;\;\;\;\;
\label{ham}
\end{eqnarray}
where $\mathbb{1}$ is the $2 \times 2$ unit matrix, $\hat{\sigma}_z$ is the third Pauli matrix,
\begin{eqnarray}
\hat{G}(\vec{r}) = -\frac{6\pi}{k_0} {\hat d}_{eg} \hat{{\mathcal G}}(\vec{r}) {\hat d}_{eg}^{\dagger},\;\;
\hat{{\mathcal G}}(\vec{r}) = -\frac{e^{i k_0 r}}{4 \pi r}
\left[ P(i k_0 r) \mathbb{1}
+ Q(ik_0 r) \frac{\vec{r} \otimes \vec{r}}{r^2} \right]
\label{greena}
\end{eqnarray}
Here $k_0 = \omega_0/c = 2\pi/\lambda_0$, $\omega_0 = (\omega_A + \omega_B)/2$,  $P(u) = 1 - 1/u + 1/u^2$, $Q(u) = -1 + 3/u - 3/u^2$, and
\begin{eqnarray}
{\hat d}_{eg} = \frac{1}{\sqrt{2}}
\begin{bmatrix}
1 & i \\
-1 & i \\
\end{bmatrix}
\label{dmatrix}
\end{eqnarray}

For $m \ne n$, a $4 \times 4$ block $\hat{H}_{mn}$ of the Hamiltonian $\hat{H}$ describes the interaction between atoms in two different unit cells $m$ and $n$. More precisely, the upper left diagonal $2 \times 2$ block of $\hat{H}_{mn}$ describes the interaction between atoms $A$ and the lower right diagonal $2 \times 2$ block---the interaction between atoms $B$. Off-diagonal $2 \times 2$ blocks of $\hat{H}_{mn}$ describe the interaction between atoms $A$ and $B$. For $m = n$, $\hat{H}_{mm}$ describes the interaction between atoms $A$ and $B$ belonging to the same unit cell $m$ (off-diagonal $2 \times 2$ blocks) and the evolution of isolated atoms (diagonal $2 \times 2$ blocks). The terms proportional to $\Delta_{AB}$ and $\Delta_{\vec{B}}$ in Eq.\ (\ref{ham}) account for the difference between resonance frequencies of atoms $A$ and $B$ and for the external magnetic field $\vec{B}$, respectively. This model is an extension of the one that was previously used to study the impact of disorder on the photonic band structure of atomic lattices \cite{antezza13} and Anderson localization of light near their band edges \cite{skip20}. 

Useful information about the behavior of the atomic lattice can be obtained from the eigenvalues  {$\Lambda_{\alpha}$} and  
 {right and left} eigenvectors $|R_{\alpha} \rangle$, $|L_{\alpha} \rangle$ of $\hat{H}$ obeying 
\begin{eqnarray}
{\hat H} | R_{\alpha} \rangle  = \Lambda_{\alpha} | R_{\alpha} \rangle, \;\;
\langle L_{\alpha} | {\hat H} = \langle L_{\alpha} | \Lambda_{\alpha}
\label{eigenrl}
\end{eqnarray}
``Quasimodes'' $|R_{\alpha} \rangle$ and $|L_{\alpha} \rangle$ can be normalized to obey $ \langle L_{\alpha} | R_{\beta} \rangle = \delta_{\alpha \beta}$ and thus constitute a biorthogonal basis over which the dynamics of the atomic system can be expanded. 
$\omega_{\alpha} = \omega_0 - (\Gamma_0/2)\text{Re} \Lambda_{\alpha}$ and $\Gamma_{\alpha} =  \Gamma_0 \text{Im} \Lambda_{\alpha} \ll \omega_{\alpha}$ are frequencies and decay rates of the quasimodes, respectively.
 {Imaginary parts of $\Lambda_{\alpha}$ arise because the considered atomic system is open and energy of atomic excitations can be emitted in the 3D space surrounding the 2D atomic lattice in the form of freely propagating electromagnetic waves.}   


The main difference between our model and the majority of other models used to study topological phenomena in condensed matter physics is the coupling between {\it all} atoms and not only between first- or second-nearest neighbors.  {This, in particular, complicates application of periodic boundary conditions when studying disordered systems of finite size as we will discuss below in Sec. {\ref{topo}}. In addition, the non-Hermiticity and the long-range nature of coupling ${\hat H}_{mn} \sim \exp{(i k_0 r_{mn})}/(k_0 r_{mn})$ may give rise to difficulties in classifying the Hamiltonian ({\ref{fullh}}) in a particular symmetry class \mbox{\cite{kawabata19, lepori17}}.}


\section{Infinite honeycomb lattice without disorder}
\label{infinite}

\begin{figure}[tbp]
\includegraphics[width=\textwidth]{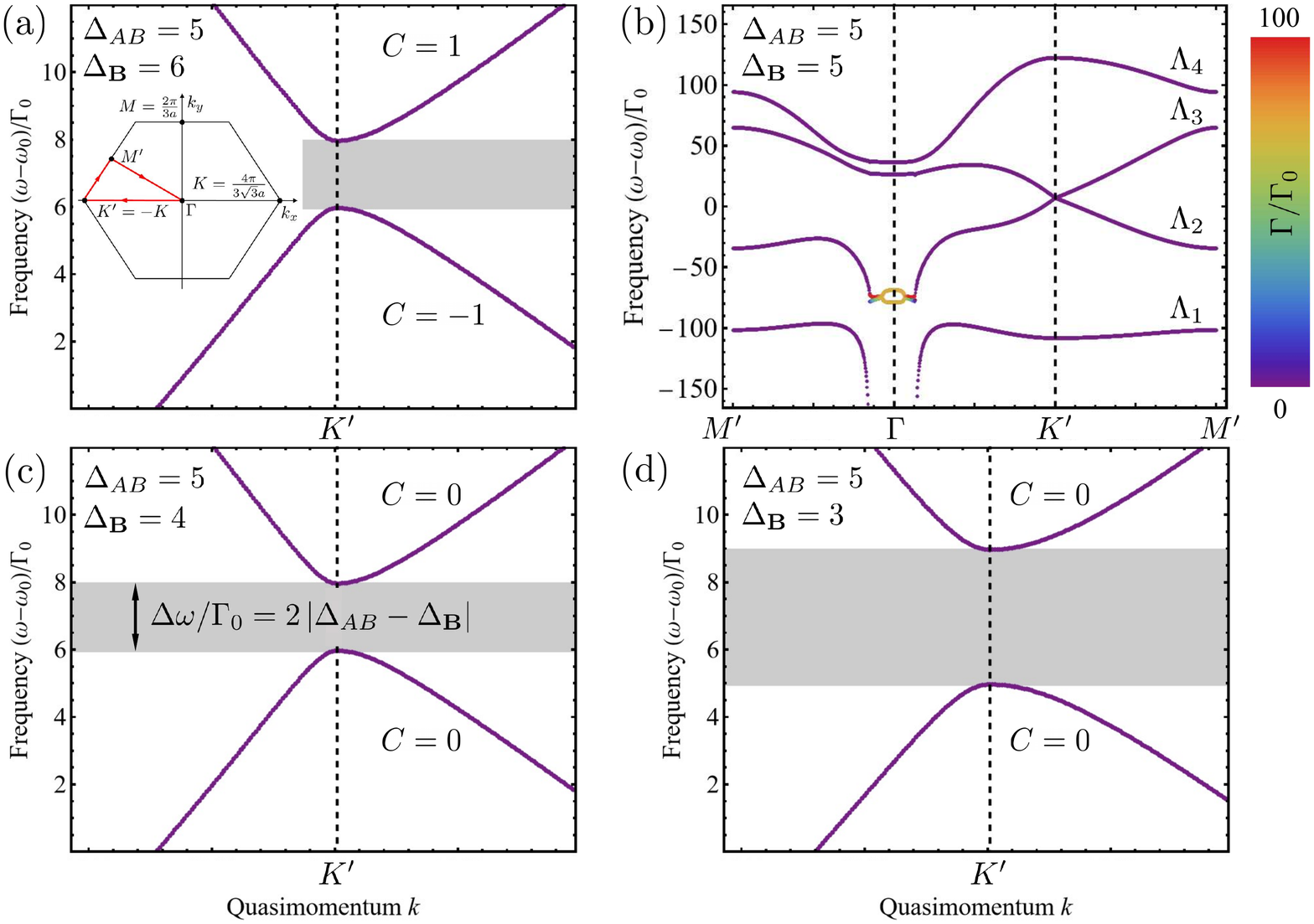}
\caption{\label{figbands}
 {Band diagram of the infinite honeycomb atomic lattice with interatomic spacing $a = \lambda_0/20$. Color code corresponds to the decay rate $\Gamma$ of quasimodes (in units of $\Gamma_0$). Vertical dashed lines indicate the positions of $\Gamma$ and $K'$ points of the Brillouin zone. The path $ M' \to \Gamma \to K' \to M'$ followed in the 2D Brillouin zone is shown in the inset of panel (a). Panel (b) shows the full band diagram for $\Delta_{\vec{B}} = \Delta_{AB} = 5$, with a crossing of two bands at $K'$ point. Panels (a), (c) and (d) zoom on the band diagram around $K'$ point to evidence the opening of a gap (shaded in gray) of width $\Delta \omega = 2\Gamma_0 |\Delta_{AB}-\Delta_{\vec{B}}|$ for $\Delta_{\vec{B}} \ne \Delta_{AB}$, the rest of the band diagram remaining very similar to the one in panel (b). Chern numbers $C$ calculated for bands below and above the gap are given in the figures. They show that the gap is topological when $|\Delta_{\vec{B}}| > |\Delta_{AB}|$ and trivial otherwise.}           
} 
\end{figure}

For the infinite lattice without disorder ($N \to \infty$, $\vec{a}_m = \vec{a}$), we can work in Fourier space. Hamiltonian becomes a $4 \times 4$ matrix $\hat{\mathcal H}(\vec{k})$  {with} eigevanues  {$\Lambda(\vec{k}) = -2 \left[ \omega(\vec{k})-\omega_0 \right]/\Gamma_0 + i \Gamma(\vec{k})/\Gamma_0$}.  
 {When $a \ll \lambda_0$, the imaginary part of $\Lambda(\vec{k})$ is different from zero only in a small part of the spectrum around $\Gamma$ point of the Brillouin zone, for $|\vec{k}| \leq k_0$, allowing for emission of electromagnetic waves out of the atomic plane, as illustrated by Fig. {\ref{figbands}}(b). We thus ignore $\Gamma(\vec{k})$ and analyze $\omega(\vec{k})$. The latter} 
form four bands,  {see Fig. {\ref{figbands}}(b)}. For small lattice spacing $a \lesssim \lambda_0/10$ and moderate $|\Delta_{\vec{B}}|$, $|\Delta_{AB}| \lesssim 10$, two of these four bands approach each other at $K$ or $K'$ points of the Brillouin zone. Depending on whether $\Delta_{\vec{B}}$ and $\Delta_{AB}$ have the same signs, these bands touch either at $K$ or $K'$ points when $|\Delta_{\vec{B}}| = |\Delta_{AB}|$ \cite{perczel17,skip22,wulles22}. Once this equality starts to be violated, a  {direct} gap opens between the two bands. 
 {Remembering that the Hamiltonian is non-Hermitian and that the spectrum is complex, we note that this is a line gap according to the terminology of Ref. \mbox{\cite{kawabata19}}: the complex eigenvalues split in two groups separated by a reference line $\Lambda = -2 (\omega_c-\omega_0)/\Gamma_0$, with $(\omega_c-\omega_0)/\Gamma_0 \simeq 7$ in Fig. {\ref{figbands}}.}
When $\Delta_{\vec{B}}$ and $\Delta_{AB}$ have the same signs, the width of the gap is controlled by $\Lambda(\vec{k})$ at $K'$ point $\vec{k} = \{ -K, 0 \}$ ($K = 4 \pi/3 \sqrt{3} a$), where $\hat{\mathcal H}(\vec{k})$ takes the following form \cite{wulles22}: 
\begin{eqnarray}
{\hat{\mathcal H}} = 
\begin{pmatrix}
{\mathcal H}_{11} & 0 & 0 & {\mathcal H}_{14}\\
0 & {\mathcal H}_{22} & 0 & 0\\
0 & 0 & {\mathcal H}_{33} & 0\\
{\mathcal H}_{41} & 0 & 0 & {\mathcal H}_{44}
\end{pmatrix}
\label{hamfourier}
\end{eqnarray} 
with
\begin{eqnarray}
{\mathcal H}_{11} &=& s+2(\Delta_{AB}+\Delta_{\vec{B}}),\;\;
{\mathcal H}_{22} = s+2(\Delta_{AB}-\Delta_{\vec{B}}),\;\;
{\mathcal H}_{33} = s + 2(-\Delta_{AB}+\Delta_{\vec{B}})
\\
{\mathcal H}_{44} &=& s+2(-\Delta_{AB}-\Delta_{\vec{B}}),\;\;
{\mathcal H}_{14} = {\mathcal H}_{41} = p
\label{hamelements}
\end{eqnarray}
The values of $s=-2 (\omega_c-\omega_0)/\Gamma_0$ and $p$ depend on the lattice spacing $a$. For $a = \lambda_0/20$, for example, we find $s \simeq -13.9$, $p = 229.8 \gg |s|$, $|\Delta_{AB}|$, $|\Delta_{\vec{B}}|$. The latter relation between $p$, $s$, $\Delta_{AB}$ and $\Delta_{\vec{B}}$ holds for other values of $a$ as well.  

The eigenvalues of ${\hat{\mathcal H}}$ are
\begin{eqnarray}
\Lambda_{1,4} &=& \frac12 \left( {\mathcal H}_{11} + {\mathcal H}_{44} \mp \sqrt{({\mathcal H}_{11}-{\mathcal H}_{44})^2 + 4 {\mathcal H}_{14} {\mathcal H}_{41}} \right)
= s \mp \sqrt{4(\Delta_{AB}+\Delta_{\vec{B}})^2 + p^2}
\simeq s \mp p
\label{ev14}
\\
\Lambda_{2} &=& {\mathcal H}_{22} = s+2(\Delta_{AB}-\Delta_{\vec{B}}),\;\;
\Lambda_{3} = {\mathcal H}_{33} = s + 2(-\Delta_{AB}+\Delta_{\vec{B}})
\label{ev23}
\end{eqnarray}
The eigenvalues $\Lambda_{1,4}$ are very different from  each other and from $\Lambda_{2,3}$ at $K'$ point because $p$ is large,  {see Fig. {\ref{figbands}}}. In contrast,  $\Lambda_{2}$ and $\Lambda_{3}$ have similar values  and coincide for $\Delta_{\vec{B}} = \Delta_{AB}$. When $\Delta_{\vec{B}} \ne \Delta_{AB}$ but have the same sign, a gap opens between the second and third eigevalues:
\begin{eqnarray}
\Lambda_{2} - \Lambda_3 &=& 4(\Delta_{AB}-\Delta_{\vec{B}})
\label{gap}
\end{eqnarray}
The middle of the gap is
\begin{eqnarray}
\frac12 \left(\Lambda_{2} + \Lambda_3 \right) &=& s
= -2 \frac{\omega_c-\omega_0}{\Gamma_0}
\label{gapmiddle}
\end{eqnarray}
The corresponding gap in frequencies of excitations is centered around $\omega_c = \omega_0 - \Gamma_0 s/2$ and has a width $\Delta \omega = 2\Gamma_0 |\Delta_{AB}-\Delta_{\vec{B}}|$.

Topological properties of wave excitations in periodic 2D lattices are commonly characterized by a topological invariant known as Chern number $C$ (see, e.g., Refs.\ \cite{cooper19} and \cite{bernevig13} for definitions). We evaluated $C$ for our system using the approach developed in Ref.\ \cite{fukui05} and found that the gap between the bands formed by the eigenvalues $\Lambda_2(\vec{k})$ and $\Lambda_3(\vec{k})$ when $\vec{k}$ varies, is topological (the sums of $C$ of bands below or above the gap are different from 0) when $|\Delta_{\vec{B}}| > |\Delta_{AB}|$ and trivial (the sums of $C$ of bands below or above the gap are both equal to 0) otherwise. This property survives a certain amount of disorder and, in particular, a topological gap preserves its nontrivial topology until the disorder is strong enough to close it \cite{skip22}. Below we show that when the gap is trivial in the absence of disorder because $|\Delta_{\vec{B}}| < |\Delta_{AB}|$, it is still possible to induce nontrivial topological properties by introducing disorder in the atomic positions.

 {The long-range nature of coupling between atoms in Eq. ({\ref{fullh}}) may lead to discontinuities in $\hat{\mathcal H}(\vec{k})$, which would complicate the analysis of its topological properties \mbox{\cite{lepori17}}. Our Hamiltonian ({\ref{hamfourier}}) indeed exhibits divergences at $|\vec{k}| = k_0$ but these are not due to the long-range nature of coupling. The divergences can be traced back to the tacit assumption of instantaneous propagation of excitations through the atomic array in Eq. ({\ref{greena}}). This assumption is justified when the size $L$ of the atomic array is smaller than $c/\Gamma_0$ but leads to unphysical divergences in $\hat{\mathcal H}(\vec{k})$ describing the infinite lattice ($L \to \infty$), as explained in Ref.  \mbox{\cite{bettles17}}. The divergences can be regularized by restoring the finiteness of the speed of propagation of excitations and introducing an additional time scale $L/c$, which, however, complicates the problem considerably. Luckily enough, the calculation of $C$ using the method of Ref. \mbox{\cite{fukui05}} can be realized even with a formally divergent $\hat{\mathcal H}(\vec{k})$ provided that the coarse mesh of $\vec{k}$ in the Brillouin zone avoids the points where $\hat{\mathcal H}(\vec{k})$ diverges.  
}

\section{Topological properties of light in a disordered lattice}
\label{topo}

 {Using} the standard definition of Chern number in terms of eigenfunctions $|\psi(\vec{k}) \rangle$ of the Hamiltonian $\hat{\mathcal H}(\vec{k})$ in a disordered system  {requires imposing twisted boundary conditions \mbox{\cite{niu84,essin07}}, which is not only quite expensive computationally, but also not straightforward in our system with long-range coupling between atoms. Indeed, periodic boundary conditions amount to putting the system on a torus, which  would reguire a substantial modification of the Green's function ${\hat {\mathcal G}}(\vec{r})$ in Eq. ({\ref{greena}}) and is likely to significantly alter system's behavior.} Various extensions of the standard definition  {of $C$} have been proposed to allow for its use in disordered lattices that can also be of finite size, including real-space representations of $C$ \cite{prodan10,prodan11,bianco11} and a recent proposal of a local invariant based on the system's spectral localizer \cite{cerjan22}. Here we use yet another topological invariant---the so-called Bott index $C_B$ \cite{loring10}.  {Bott index has been used extensively in recent years to characterize topological properties of photonic systems that lack periodicity \mbox{\cite{bandres16,stutzer18,lustig19,liu20,zhou20}}.} In a rectangular sample of sides $L_x$ and $L_y$, it is given by \cite{loring10}
\begin{eqnarray}
C_B(\omega) = \frac{1}{2\pi} \mathrm{Im} \mathrm{Tr} \ln \left[ {\hat V}_X(\omega) {\hat V}_Y(\omega) {\hat V}_X^{\dagger}(\omega) {\hat V}_Y^{\dagger}(\omega)  \right]
\label{bott}
\end{eqnarray}
where
${\hat V}_{X,Y}(\omega)  = {\hat P}(\omega) {\hat U}_{X,Y} {\hat P}(\omega)$, 
${\hat U}_X = \exp(i 2\pi {\hat X}/L_x)$,
${\hat U}_Y = \exp(i 2\pi {\hat Y}/L_y)$ and
\begin{eqnarray}
\hat{X} =
\begin{bmatrix}
x_1 & 0 &  \ldots  & 0 & 0 \\
0 & x_2 &  \ldots  & 0 & 0 \\
\ldots & \ldots  & \ldots & \ldots & \ldots \\
0 & 0 &  \ldots &  x_{N-1} & 0 \\
0 & 0 &  \ldots &  0 & x_N \\
\end{bmatrix},\;\;
\hat{Y} =
\begin{bmatrix}
y_1 & 0 & \ldots & 0 & 0 \\
0 & y_2 & \ldots & 0 & 0 \\
\ldots & \ldots &  \ldots & \ldots & \ldots \\
0 & 0 & \ldots & y_{N-1} & 0 \\
0 & 0 & \ldots & 0 & y_N \\
\end{bmatrix}
\label{xy}
\end{eqnarray}
A projector operator on quasimodes corresponding to frequencies $\omega_{\alpha}$ below $\omega$ is defined as \cite{song19,skip22}
\begin{eqnarray}
{\hat P}(\omega) = \sum\limits_{\omega_{\alpha} \leq \omega} | R_{\alpha} \rangle \langle L_{\alpha} |
\label{proj}
\end{eqnarray}

\begin{figure}[tbp]
\includegraphics[width=\textwidth]{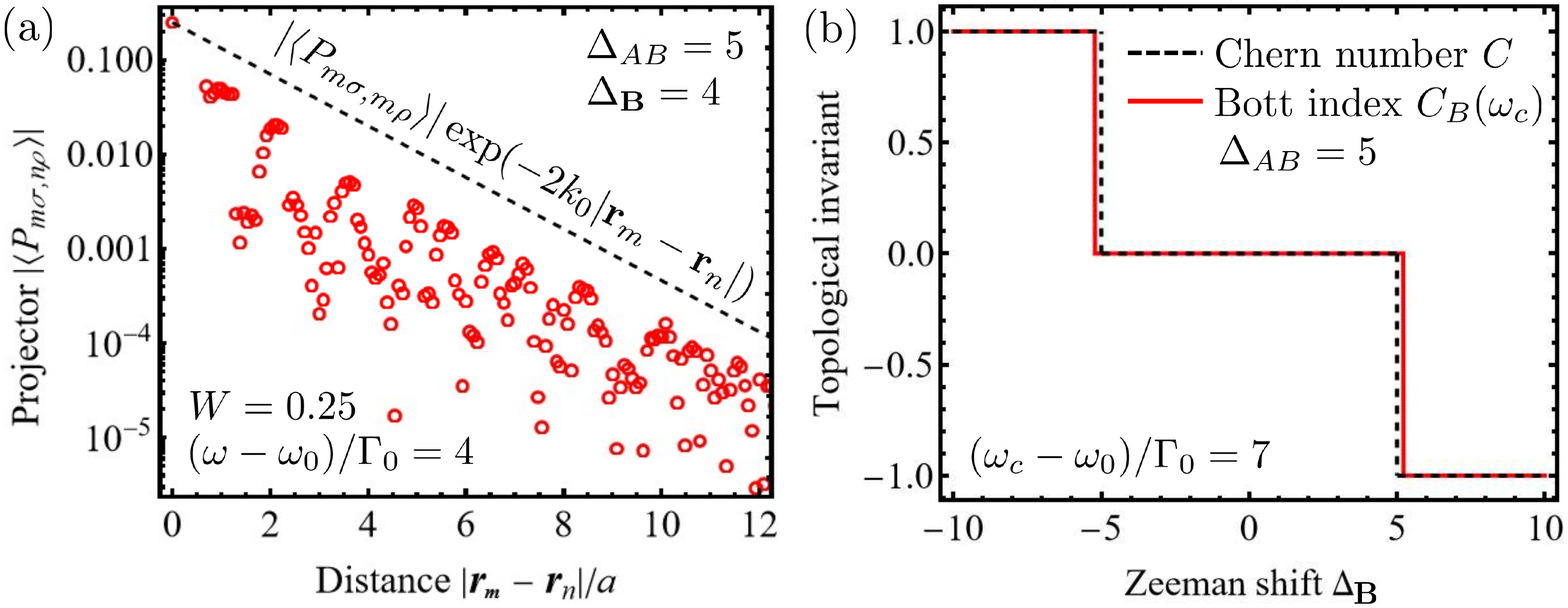}\vspace{-4.5cm}
\caption{\label{figtest}
 {(a) Absolute value of the average matrix element $P_{m \sigma, n \rho}$ of the projector ({\ref{proj}}) as a function of the 
distance between atoms $m$ and $n$. Parameters are chosen in the middle of the TAI phase (violet `island') in Fig. {\ref{figbott}}(c). Averaging is performed over the two polarization components $\sigma, \rho = \pm 1$ and over 10 independent realizations of disorder. Dashed line illustrates the exponential decay of $| \langle P_{m \sigma, n \rho} \rangle |$.
(b) Chern number $C$ of the two lower bands and Bott index $C_B(\omega_c)$ for $(\omega_c - \omega_0)/\Gamma_0 = 7$ in the middle of the band gap between the second and third bands  (see Fig. {\ref{figbands}}), as functions of $\Delta_{\vec{B}}$ for a fixed $\Delta_{AB} = 5$. $C_B$ is calculated for a square sample as in Fig. {\ref{figlattices}}(a) made of $N = 2244$ atoms and without disorder ($W = 0$).
}}
\end{figure}

 {Bott index defined by Eqs. ({\ref{bott}}--{\ref{proj}}) has proved to be an efficient tool for characterization of topological properties of non-Hermitian systems with open boundaries \mbox{\cite{zeng20prb,tang20}}. The non-Hermitian ${\hat P}(\omega)$ defined by Eq. ({\ref{proj}}) instead of the standard Hermitian
${\hat P}_{\text{Herm}}(\omega) = \sum_{\omega_{\alpha} \leq \omega} | R_{\alpha} \rangle \langle R_{\alpha} |$
ensures integer values of $C_B$ and suppresses its fluctuations from one realization of disorder to another, allowing for faster convergence of averaging over disorder. Apparently, using the definition ({\ref{proj}}) allows for partially mitigating the effect of non-Hermiticity of the Hamiltonian by preserving the idempotence of ${\hat P}(\omega)$.
For the Bott index ({\ref{bott}}) to make sense with the definition ({\ref{proj}}) of ${\hat P}(\omega)$, we have to ensure that the latter is local \mbox{\cite{loring10}}. To this end, we compute the average of the matrix element $P_{m \sigma, n \rho}$ over polarizations $\sigma, \rho = \pm 1$ and over disorder for typical parameters used in our further calculations. The result is shown in Fig. {\ref{figtest}}(a) as a function of the distance between atoms $m$ and $n$. $| \langle P_{m \sigma, n \rho} \rangle |$ decays exponentially with the distance, which validates the use Eq. ({\ref{proj}}). 
}

 {Although the equivalence of Bott index $C_B$ and Chern number $C$ has been proven \mbox{\cite{toniolo22}}, our Hamiltonian ${\hat H}$ does not satisfy some of the conditions required by the proof. In particular, ${\hat H}$ is non-Hermitian, long-range, and describes an open atomic lattice (because of both the open boundaries in the plane of the lattice and the leakage of energy out of the plane), as we discussed above. It is thus worthwhile to compare $C_B$ and $C$ explicitly for an atomic lattice without disorder  ($\vec{a}_m = \vec{a}$) where both can be evaluated. We perform such a comparison in Fig. {\ref{figtest}}(b) and observe excellent agreement. $C_B$ slightly overestimates the width of the topologically trivial region where $C = 0$ but the difference between $C_B$ and $C$ reduces when increasing the number of atoms $N$, and we expect it to vanish in the thermodynamic limit $N \to \infty$.        
}

\begin{figure}[tbp]
\includegraphics[width=\textwidth]{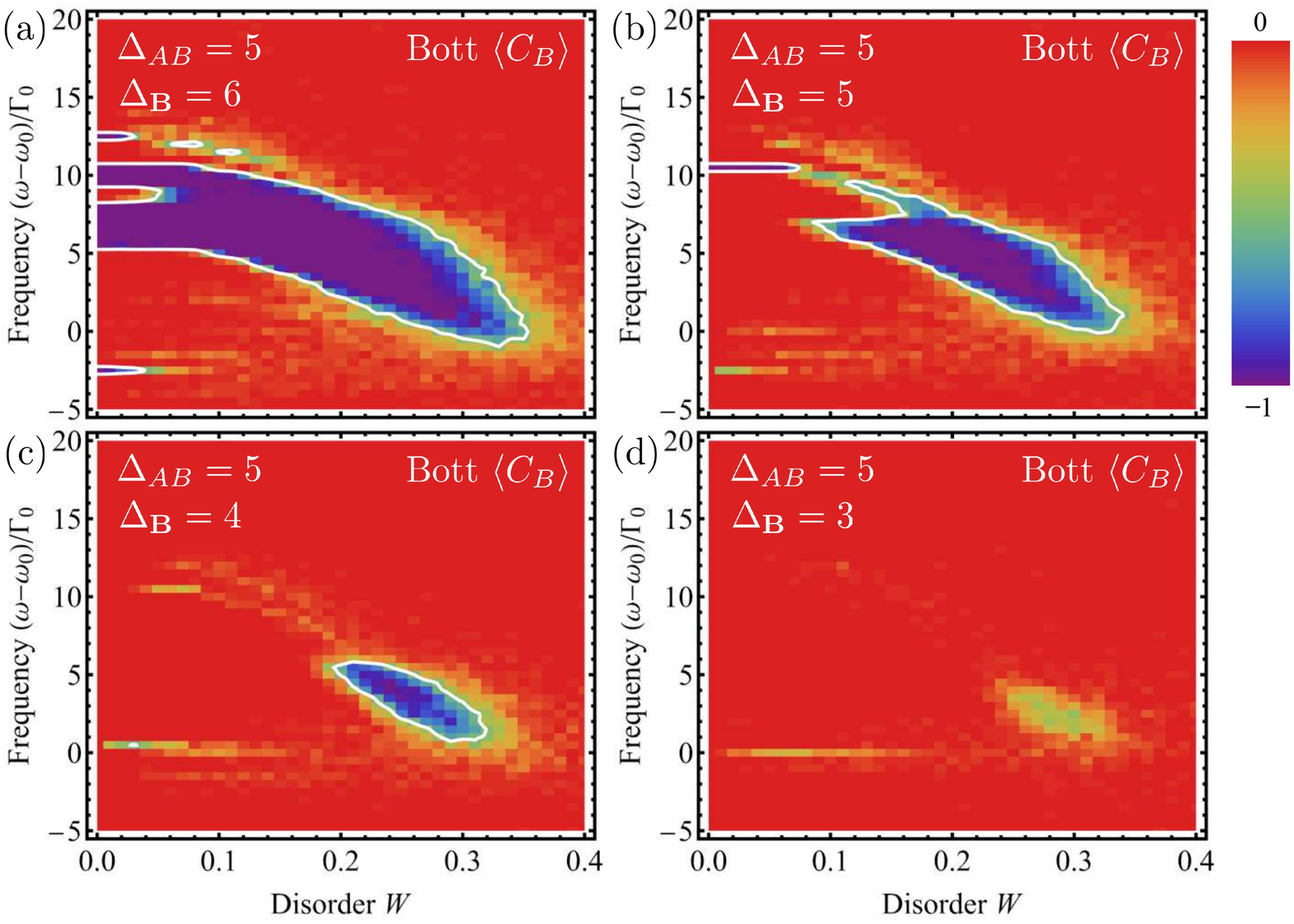}
\caption{\label{figbott}
Average Bott index $\langle C_B \rangle$ calculated for four different values of $\Delta_{\vec{B}}$ around $\Delta_{\vec{B}} = \Delta_{AB} = 5$ for $a = \lambda_0/20$ in a rectangular sample shown in Fig.\ \ref{figlattices}(a). The white contour shows the boundary $\langle C_B \rangle = -0.5$ between the topologically nontrivial region where $\langle C_B \rangle = -1$ (violet) and the trivial region where $\langle C_B \rangle = 0$ (red). Ensemble averaging is performed over 60 independent realizations of disorder for each graph.   
} 
\end{figure}

 {Armed with a reliable tool to characterize the topological properties of our photonic system, we turn to a study of the impact of disorder.}
Figure \ref{figbott} shows the average Bott index in a disordered  {atomic} lattice for four values of $\Delta_{\vec{B}}$ around $\Delta_{\vec{B}} = \Delta_{AB} = 5$. When  $\Delta_{\vec{B}} > \Delta_{AB}$, the lattice exhibits a topological band gap already in the absence of disorder, which yields  $\langle C_B \rangle \ne 0$ in a narrow band of frequencies on the vertical axis $W = 0$, see Figs.\  {{\ref{figbands}}(a) and} \ref{figbott}(a). The impact of disorder is to reduce the width of the gap and to close it eventually, as we discussed in detail in Ref.\ \cite{skip22}. 
Notice that the gap shifts to lower frequencies with disorder but its width is hardly affected up to $W \simeq 0.3$, which can be seen as a sign of robustness of the topological band gap to disorder.
 {In this regime, our system is a disordered TI.}

At a smaller $\Delta_{\vec{B}} = \Delta_{AB}$, there is no band gap for $W = 0$ 
 {as we show in Fig. {\ref{figbands}}(b)} 
but Bott index signals topologically nontrivial properties to arise starting from $W \simeq 0.1$, see Fig.\ \ref{figbott}(b). These topological properties are {\it induced} by disorder and are therefore a signature of TAI. They persist up to a relatively large disorder $W \simeq 0.35$, after which the spectrum becomes trivial again. It may appear surprising that the topologically nontrivial region does not start immediately once $W \ne 0$ and require a threshold value of $W$ to develop already for $\Delta_{\vec{B}} = \Delta_{AB}$. However, this may be an artifact of finite sample size in our calculations. In a finite sample, the spectrum is discrete and a band gap or any frequency-dependent property can be identified robustly only on scales superiour to mode spacing. The region with nontrivial topological properties in Fig.\ \ref{figbott}(b) opens near $(\omega-\omega_0)/\Gamma_0 = -s/2 \simeq 7$ where the density of states (DOS) is particularly low and hence the mode spacing is large. In fact, $\text{DOS} = 0$ in the absence of disorder ($W = 0$) at exactly $(\omega-\omega_0)/\Gamma_0 = -s/2$ and $\Delta_{\vec{B}} = \Delta_{AB}$.

Figure \ref{figbott}(c) illustrates a less ambiguous situation of $\Delta_{\vec{B}} < \Delta_{AB}$. In this case, there is a topologically trivial band gap centered at $(\omega-\omega_0)/\Gamma_0 \simeq 7$ and having a width $\Delta \omega/\Gamma_0 = 2$ on the vertical axis $W = 0$
 {[see Fig. {\ref{figbands}}(c)]}
but, as expected, it is not visible in the Bott index. The latter becomes different from zero for $W$ between roughly 0.2 and 0.32 signaling TAI phase in this range of disorder strengths. Finally, when the difference between $\Delta_{\vec{B}}$ and $\Delta_{AB}$ increases further, TAI becomes less pronounced. For the parameters of Fig.\ \ref{figbott}(d), for example, $C_B$ is found to be significantly different from zero only for particular realizations of disorder and its average value never decreases below $-0.5$. Thus, TAI phase only exists when $|\Delta_{\vec{B}}|$ is not too different from $|\Delta_{AB}|$.

Figures \ref{figbott}(a--d) exhibit a topologically nontrivial artifact around $(\omega-\omega_0)/\Gamma_0 \simeq 10.5$ at weak disorder. It is due to modes that arise in the corners of the rectangular atomic lattice used for the calculation of $C_B$ [see Fig.\ \ref{figlattices}(a)]. Other artifacts around $(\omega-\omega_0)/\Gamma_0 \simeq 13$, 0 and $-3$ are also  due to the finite size of the considered atomic lattice.

\section{Bulk and edge modes}
\label{dos}

\begin{figure}[tbp]
\includegraphics[width=\textwidth]{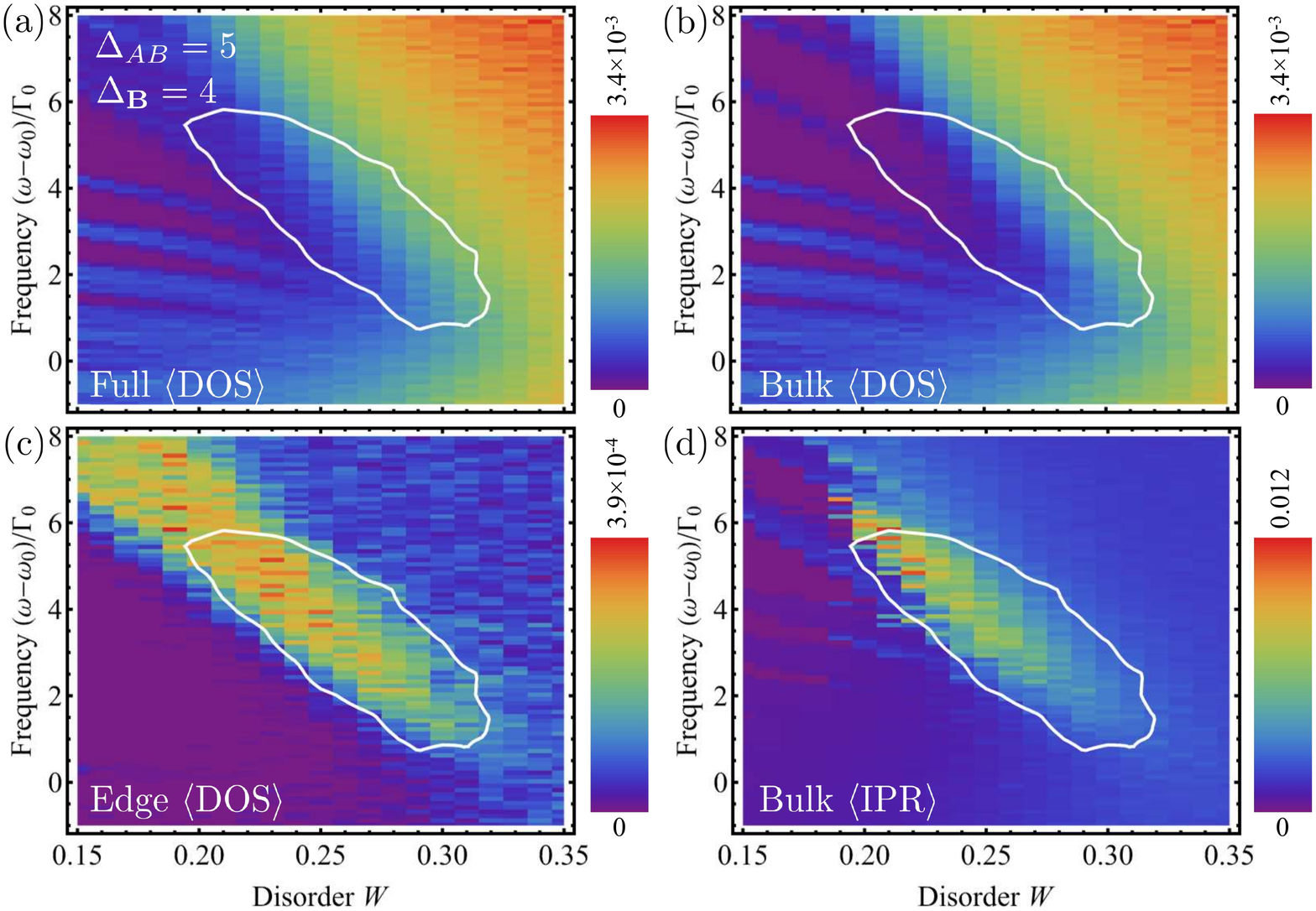}
\caption{\label{figdosipr}
Average full (a), bulk (b) and edge (c) DOS in a hexagon-shaped sample shown in Fig.\ \ref{figlattices}(b) for $a =\lambda_0/20$,  $\Delta_{AB} = 5$ and $\Delta_{\vec{B}} = 4$. Average IPR of bulk modes is shown in (d). Ensemble averaging is performed over 200 independent realizations of disorder for each graph. The white contour in (a--d) shows the boundary of the topologically nontrivial region from Fig.\ \ref{figbott}(c). 
} 
\end{figure}


The key property of TAI is their ability to host propagating states at the edges while being insulating in the bulk. It is important for experimental observation of TAI as well as for their potential applications. Let us explore how this property is realized for optical modes of our cold-atom lattice. We focus on the case of $\Delta_{AB} = 5$, $\Delta_{\vec{B}} = 4$ for which the Bott index is shown in Fig.\ \ref{figbott}(c). To examine phenomena due to edge modes separately from those associated with the bulk of the sample, we separate a hexagon-shaped sample in bulk and edge parts as illustrated in Fig.\ \ref{figlattices}(b). Quasimodes $| R_{\alpha}    \rangle$ that have more than 50\% of their weight $\langle R_{\alpha}| R_{\alpha} \rangle$ inside the bulk part of the sample are considered as ``bulk modes'' whereas quasimodes with more than 50\% of their weight coming from the edge of the sample are called ``edge modes''.

Figures \ref{figdosipr}(a), (b) and (c) show the average DOS of all, bulk and edge modes, respectively. We see that, clearly, the topologically nontrivial region delimited by the white line in Figs.\ \ref{figdosipr}(a--d) does not correspond to a gap in the full DOS shown in Fig.\ \ref{figdosipr}(a). Bulk DOS in Fig.\ \ref{figdosipr}(b) exhibits some reduction with respect to Fig.\ \ref{figdosipr}(a) 
 {but no gap opens in the spectrum.}
The result for the average DOS of edge states in Figs.\ \ref{figdosipr}(c)
 {clearly indicates that} 
for $W$ between 0.2 and 0.32, the frequencies of edge states are concentrated inside the topologically nontrivial region where $\langle C_B \rangle = -1$. Thus, the edges of the sample indeed host edge modes when  $\langle C_B \rangle \ne 0$.

To understand the nature of bulk modes in the TAI phase, we compute their inverse participation ratio (IPR),
\begin{eqnarray}
\text{IPR}_{\alpha} &=& \frac{\sum_{m=1}^N \left( \sum_{\sigma = \pm 1} |R_{\alpha m \sigma}|^2 \right)^2}{\left( \sum_{m=1}^N \sum_{\sigma = \pm 1} |R_{\alpha m \sigma}|^2 \right)^2}
\label{ipr}
\end{eqnarray}   
where $R_{\alpha m \sigma}$ is the amplitude of the $\sigma$-polarized component of quasimode $| R_{\alpha} \rangle$ on the atom $m$. In the notation introduced in Sec.\ \ref{model}, all components of $| R_{\alpha} \rangle$ are organized in a vector of length $2N$, so that $R_{\alpha m \sigma}$ is the $[2m + (\sigma-1)/2]$-th component of the vector $| R_{\alpha} \rangle$:  $R_{\alpha m \sigma} = (R_{\alpha})_{2m + (\sigma-1)/2}$.
IPR defined by Eq.\ (\ref{ipr}) quantifies the spatial localization of the quasimode $| R_{\alpha} \rangle$ with account for both polarization states. For a mode localized on a single atom, $\text{IPR} \sim 1$ whereas for a mode extended over $M$ atoms  $\text{IPR} \sim 1/M$. An estimate of the localization length $\xi$ of a mode can be obtained from its $\text{IPR}$ as $\xi \sim a \times \text{IPR}^{-1/2}$.  

The average IPR of bulk modes is shown in Fig.\ \ref{figdosipr}(d). We see that the modes with frequencies inside the topologically nontrivial region of parameter space have a tendency to have larger IPR than the modes outside this region.
 {This observation is consistent with the fact that the topologically nontrivial region corresponds to a mobility gap, which is a distinctive property of TAI \mbox{\cite{li09,prodan11,zhou20}}.}
Some of the modes with large IPR arise in the topologically trivial region as well, likely because of inaccuracies in determination of the precise location of the phase boundary due to finite-size effects. Simultaneous analysis of the four graphs of Fig.\ \ref{figdosipr} suggests that our results exhibit all the properties characteristic of TAI: a topologically nontrivial region with a topological invariant  $C_B \ne 0$ arises due to disorder and favors the appearance of edge modes and the spatial localization of bulk modes. Thus, an optical transport experiment is expected to show propagation of optical energy along edges and bad optical conductance of the bulk, as expected for TAI.

 {Our approach to distinguishing bulk and edge modes is not perfect because some of the strongly localized modes that do not require a sample edge for their existence, may accidentally arise close to an edge and be identified as edge modes. Such modes then contribute to edge DOS in Fig. {\ref{figdosipr}}(c). However, there is no reason for strongly localized modes to concentrate near edges and they are likely to be distributed uniformly over the sample area. Thus, although edge DOS in Fig. {\ref{figdosipr}}(c) can be contaminated by localized states, the difference between edge DOS in Fig. {\ref{figdosipr}}(c) and bulk DOS Fig.  {\ref{figdosipr}}(b) is due to genuine edge modes only. The comparison of  Figs. {\ref{figdosipr}}(c) and (b) shows that edge DOS is enhanced mainly in the spectral regions where bulk DOS is suppressed, confirming the ability of our approach to distinguishing bulk and edge modes. Note also that in small lattices where the edge region constitutes a considerable part of the system [as in the illustration of Fig. {\ref{figlattices}}(b), for example] some of weakly localized or extended modes can be mistaken for edge ones. However, the probability of this decreases with the increase of lattice size (at a constant $h_{\text{edge}}$) and becomes negligible for lattices for which our calculations are performed and which contain almost 20 times more atoms than the lattice in Fig. {\ref{figlattices}}(b).}

\section{Discussion}
\label{discuss}     

\begin{figure}[tbp]
\includegraphics[width=\textwidth]{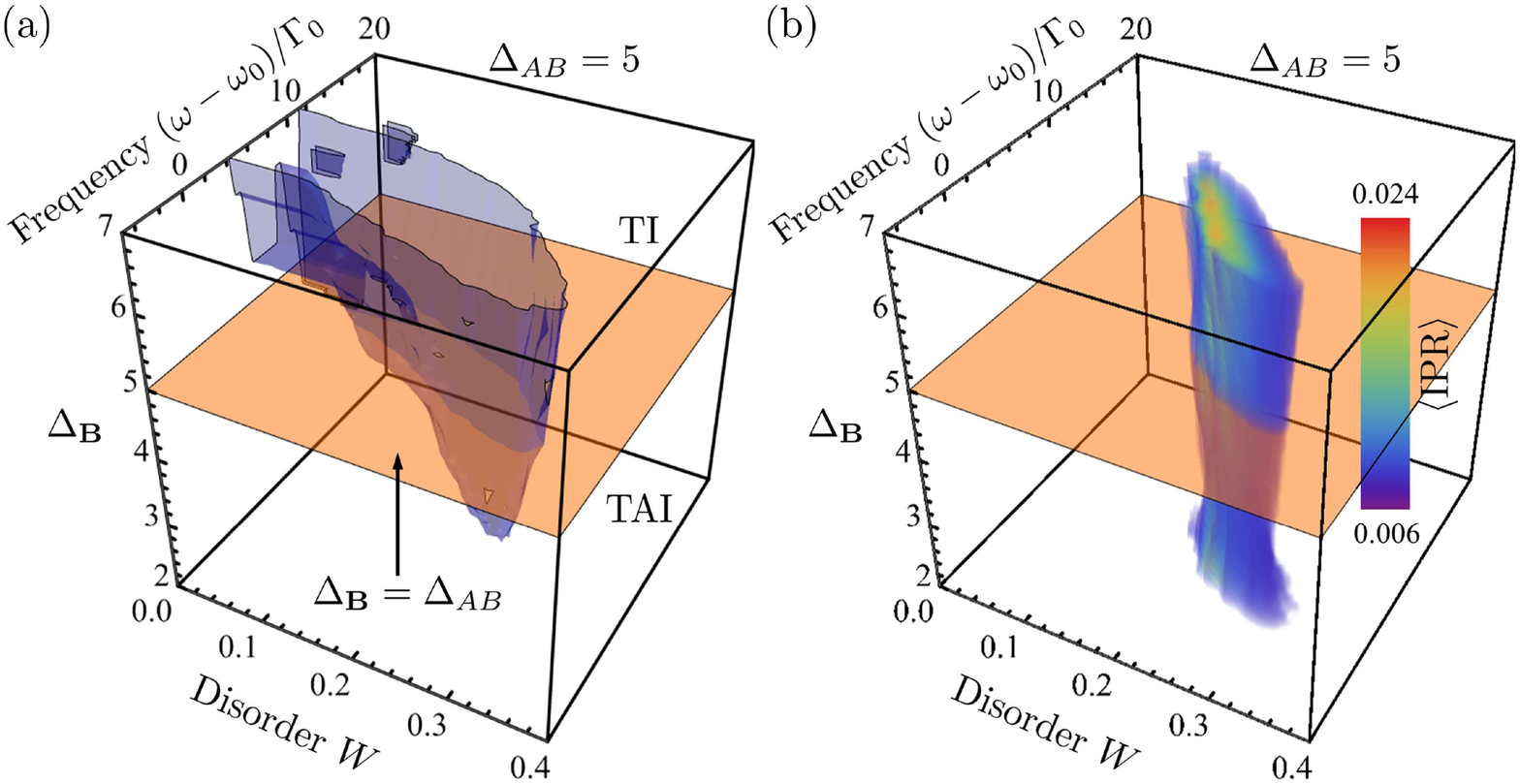}
\vspace{-3cm}
\caption{\label{fig3d}
(a) Phase diagram of the 2D disordered honeycomb atomic lattice in the 3D parameter space ($W$, $\omega$, $\Delta_{\vec{B}}$)  for $a =\lambda_0/20$ and $\Delta_{AB} = 5$. The blue surface shows the boundary between the topologically nontrivial (inside, $\langle C_B \rangle = -1$) and trivial (outside, $\langle C_B \rangle = 0$) regions of the parameter space. The orange plane $\Delta_{\vec{B}} = \Delta_{AB}$ separates the TI phase for $\Delta_{\vec{B}} > \Delta_{AB}$ from TAI phase for $\Delta_{\vec{B}} < \Delta_{AB}$.
(b) 3D density plot of average bulk IPR with the orange plane separating TI and TAI as in (a). For clarity, the values less than 0.006 are not shown (transparent). }
\end{figure}

Instead of showing the boundary $\langle C_B \rangle = -0.5$ between topologically trivial and nontrivial regions of parameters as a line in ($W$, $\omega$) plane for given $\Delta_{AB}$ and $\Delta_{\vec{B}}$ as in Fig.\ \ref{figbott}, it is instructive to show it as a surface in a three-dimensional (3D) parameter space  ($W$, $\omega$, $\Delta_{\vec{B}}$) for a given $\Delta_{AB}$.  This results in a blue surface in Fig.\ \ref{fig3d}(a). Such a representation clearly shows that one can go from TI phase for  $\Delta_{\vec{B}} > \Delta_{AB}$ (and possibly $W = 0$) to TAI phase for $\Delta_{\vec{B}} < \Delta_{AB}$ (and $W$ exceeding some threshold value depending on $\Delta_{\vec{B}}$, $\Delta_{AB}$) without changing the value of the topological invariant. Thus, no topological phase transition separates TI and TAI, which therefore correspond to the same topological phase. This is in contradiction with the (erroneous) impression that one can get from Fig.\ \ref{figbott} and with some initial expectations \cite{li09} suggesting that TAI and TI might correspond to two different topological phases. In a broader context, the fact that TAI is connected to TI and that the two represent the same topological phase has been previously established for models of QSH insulators  \cite{prodan11prb,yamakage13}.  

Figure \ref{fig3d}(b) shows the average IPR of  bulk states as a function of $W$, $\omega$ and $\Delta_{\vec{B}}$ for $\Delta_{AB} = 5$. We see that $\langle \text{IPR} \rangle$ remains consistently large inside the topologically nontrivial part of the parameter space. On the one hand, Fig.\ \ref{fig3d}(b) illustrates an important difference between TAI and TI in the clean limit: whereas a {\it clean} TI ($W = 0$) has no modes in the bulk for frequencies inside the band gap, the bulk of TAI contains spatially localized modes. On the other hand, we see that localization properties of modes in the bulk of a sufficiently {\it disordered} TI ($W \gtrsim 0.1$) are very similar to those of TAI and that the transition between TI and TAI does not lead to any significant modification of mode extent $\xi \propto \text{IPR}^{-1/2}$. 

Whereas the behavior of the topological invariant shown in Fig.\ \ref{figbott} resembles the one found previously for other topologically nontrivial systems with disorder \cite{li09,groth09,prodan10,prodan11,liu17,stutzer18,liu20}
the behavior of $\langle \text{IPR} \rangle$ and the corresponding localization properties of quasimodes in our system 
 {require special attention. We remind that} in models intended to describe electrons in 2D disordered solids \cite{li09,groth09,prodan10,prodan11} or light in a given polarization state \cite{liu17,stutzer18,liu20}, strong disorder is expected to result in spatial localization of {\it all} states apart from those near the topological phase boundary in the parameter space \cite{belissard94,prodan10,prodan11}. 
 {In particular, the states belonging to the topologically trivial part of the phase diagram are expected to be localized at strong disorder. On the contrary, light in an ensemble of atoms at random positions} exhibits Anderson localization neither in 3D \cite{skip14,tiggelen21} nor in 2D for the TE polarization considered here \cite{maximo15}.
 {Unfortunately, the system considered in the present work is somewhat intermediate between 2D and 3D and does not correspond exactly to either of atomic ensembles of Refs. \mbox{\cite{skip14,tiggelen21,maximo15}}. In addition, our results do not allow us to make a definitive conclusion about localization properties of quasimodes at strong disorder. We thus leave this question for a separate future study but note here that the peculiarities of Anderson localization of light may have an impact on} the properties of the topological phase transition taking place at the surface in parameter space shown in Fig.\ \ref{fig3d}(a). 

Finally, we comment on the possible experimental observation of phenomena reported in this work. 2D lattices of cold atoms can be realized by loading atoms in an appropriate optical interference pattern in which the atoms experience a force that pushes them towards maxima (or minima) of optical intensity \cite{soltan11,tarruell12}. However, the spacing $a$ between atoms in such lattices cannot be made less that $\lambda/2$, with $\lambda \sim \lambda_0$ the wavelength of light used to create the lattice.  The figures of this paper correspond to a much shorter spacing $a = \lambda_0/20$ and we expect our results to hold up to $a \simeq \lambda_0/10$ but not beyond \cite{skip22,wulles22}. Such subwavelength spacings between atoms require more sophisticated approaches to be realized.  {Notable examples of the latter are} schemes based on spin-dependent optical lattices with a time-periodic modulation \cite{nas15}, on resonantly Raman-coupled internal degrees of freedom \cite{anderson20},  {or on a combination of a ``magic'' wavelength with long-wavelength dipolar transitions \mbox{\cite{olmos13}}}.

The Hamiltonian considered in this work is a very good approximation for two-level atoms that are much smaller than the optical wavelength. Beyond cold-atom optics, it can also be used to predict, at least to some extent, the optical properties of lattices of dielectric resonators of finite size near geometrical (Mie) scattering resonances. Such lattices have been recently used to study topological phenomena with microwaves \cite{wang09,liu20,ma19b,ma20,reisner21}. A response to an external magnetic field can be introduced by using a gyromagnetic material (such as, e.g., YIG) for some of the components of the lattice \cite{ma20}. In particular, this approach has recently allowed to realize a microwave TAI \cite{liu20}. We believe that lattices of microwave resonators is a promising experimental platform for studies of topological phenomena with electromagnetic waves.

\section{Conclusions}
A 2D honeycomb lattice of cold atoms with subwavelength interatomic spacing is a promising candidate for realizing a photonic TAI. We use Bott index $C_B$ to characterize the topological properties of the lattice for in-plane polarization of propagating light, establish its topological phase diagram, and determine the ranges of parameters for which the system is a photonic TI or a photonic TAI. A transition between TI and TAI can take place at a constant value of $C_B$, confirming that TI and TAI correspond to the same topological phase. Our calculations show that the bulk DOS is suppressed for TAI 
 {although not enough to open a gap in the spectrum.}
The suppression of bulk DOS is accompanied by an increase of the number of edge modes and spatial localization of modes in the bulk of the lattice, 
 {leading to the opening of a mobility gap}. The interplay between disorder-induced (Anderson) localization of modes and topological phenomena in photonics
 {may present} a number of differences with respect to the better studied case of electronic or scalar-wave systems due to the  {possible} suppression of Anderson localization
 {even at strong disorder \mbox{\cite{skip14,tiggelen21}}. A detailed study of this issue remains to be performed.}      

\section{Acknowledgments}
This work was funded by the Agence Nationale de la Recherche (Grant No. ANR-20-CE30-0003 LOLITOP).

\bibliographystyle{crunsrt}


\bibliography{refstopo}

\end{document}